\documentclass[11pt,a4paper]{article}

\usepackage[margin=1in]{geometry}
\usepackage{amssymb}
\usepackage{amsmath}
\usepackage{multirow}
\usepackage{graphicx}
\usepackage{hyperref}
\usepackage{authblk}

\newcommand{\orcid}[1]{\unskip\textsuperscript{\href{https://orcid.org/#1}{[ORCID]}}}

\title{Quantum Approximate Multi-Objective Optimization in Routing Problems}

\author[1, 2, 3, *]{Eduardo Willwock Lussi\orcid{0009-0008-8805-3057}}
\author[1]{Alisson dos Passos Fumaco\orcid{0009-0003-3258-5146}}
\author[1]{Marcos Vinicius Reballo\orcid{0009-0003-6396-9302}}
\author[1, 4]{Jose Carlos Libois Neto\orcid{0000-0003-2854-378X}}
\author[1]{Fernando Augusto Caletti de Barros\orcid{0009-0008-5152-7036}}
\author[1, 4]{Eduardo Inacio Duzzioni\orcid{0000-0002-8971-2033}}

\affil[1]{Eldorado Research Institute, Porto Alegre, Brazil}
\affil[2]{Informatics and Statistics Department, Federal University of Santa Catarina, Florianópolis, Brazil}
\affil[3]{Quantuloop, Salvador, Brazil}
\affil[4]{Physics Department, Federal University of Santa Catarina, Florianópolis, Brazil}
\affil[*]{Author to whom any correspondence should be addressed: \texttt{eduardo.lussi.BE@eldorado.org.br}}

\date{}

\begin{document}

\maketitle

\begin{abstract}
Multi-objective optimization (MOO) problems are common in logistics, where routing decisions must balance conflicting objectives such as travel distance, delivery time, and operational risk. Recently, a Quantum Approximate Optimization Algorithm (QAOA) parameter-transfer strategy was proposed to efficiently solve multi-objective MAX-CUT problems by reusing parameters trained on smaller instances, avoiding the costly optimization of QAOA parameters for every scalarized problem. However, its effectiveness has only been demonstrated on proof-of-concept MAX-CUT instances specifically designed to match quantum hardware connectivity. In this work, we investigate whether this strategy remains effective for practical multi-objective routing problems. We formulate the Traveling Salesman Problem (TSP) and the Vehicle Routing Problem (VRP) as Quadratic Unconstrained Binary Optimization (QUBO) problems, reduce them to MAX-CUT, and evaluate the parameter-transfer framework under the same experimental conditions adopted in the original work. We validate this parameter transfer strategy not only through classical simulations but also via experiments on real IBM quantum hardware. The generated Pareto fronts are compared against those obtained by an adapted classical $\epsilon$-constraint method using hypervolume as the main quality indicator. Experimental results show that parameter transfer remains effective in this more realistic setting, frequently producing solutions with higher hypervolume while often reaching these solutions earlier under the adopted computational assumptions. However, its performance strongly depends on the problem structure. While the approach is consistently effective for the TSP, it becomes considerably less stable for the more constrained VRP, indicating that the transferability of QAOA parameters decreases as the underlying optimization landscape becomes more complex. These findings provide the first comprehensive evaluation of QAOA parameter transfer on realistic multi-objective routing problems and offer insights into the applicability of this strategy beyond proof-of-concept MAX-CUT benchmarks.
\vspace{1ex}
\end{abstract}

\section{Introduction}
Multi-objective optimization (MOO) problems are those that contain more than one objective function to be optimized \cite{Ehrgott2005}. An objective function is a mathematical expression that quantifies the performance or quality of a solution according to a specific criterion that must be maximized or minimized. In most cases, these functions represent conflicting objectives, where improving one is only possible if others are degraded. The solution of a MOO problem then consists of a set of solutions that represent optimal trade-offs between the objectives. The concept of optimality adopted in this context is the Pareto optimal. A solution is said to be Pareto-optimal when it is not possible to improve one objective without degrading at least one of the others.

MOO problems can be found in a wide range of applications, where multiple and often conflicting objectives must be considered simultaneously. In particular, they are highly prevalent in logistics \cite{He2023, Zhang2024, Homayouni2023, Zarejeddi2025}. This paper addresses the Vehicle Routing Problem (VRP) by investigating two distinct variants: the classical Traveling Salesman Problem (TSP) \cite{Lucas2014}[Section 7] and a more complex, custom VRP formulation adapted from recent literature \cite{vrp2021}. In both problems, the goal is to visit or deliver across a set of locations using the best possible route. The main difference between VRP and TSP is that, in VRP, multiple vehicles are available. In these problems, the definition of the ``best possible route'' depends on the objectives considered, such as minimizing distance, travel time, risk, or cost. In MOO, these criteria can be optimized simultaneously, resulting in a set of Pareto-optimal routes that represent different trade-offs between the objectives.

Many works in the literature address MOO formulations of logistics and routing problems. For example, \cite{He2023} studies a multi-objective VRP with heterogeneous vehicles and multiple distribution centers, simultaneously optimizing transportation cost, delivery time, and vehicle utilization. In \cite{Homayouni2023}, the authors investigate a green supply chain design problem involving multiple vehicle types, carbon regulation policies, and uncertainty in demand and costs, aiming to minimize both total cost and environmental impact. Similarly, \cite{Zhang2024} addresses the logistics of prefabricated construction components under a just-in-time strategy, considering carbon emissions and delivery punctuality as competing objectives under uncertainty. Finally, \cite{Zarejeddi2025} explores the design of multi-level supply chain and distribution networks using robust optimization with interval and real-time traffic data, optimizing cost, emissions, and service level.

Solving MOO problems is considerably hard due to their high combinatorial complexity and operational constraints. Classical approaches to solving MOO problems can be broadly categorized into a priori and a posteriori methods \cite{Hwang1979}. A priori methods incorporate preference information from the decision maker before optimization, often by scalarizing the multi-objective problem into a single-objective formulation, such as through weighted-sum methods, where predefined weights are assigned to the objectives according to the desired preferences. These approaches are generally more computationally efficient, but weighted-sum methods cannot recover non-supported Pareto-optimal solutions located in non-convex regions of the Pareto front. On the other hand, a posteriori methods generate a representative set of Pareto-optimal solutions before incorporating the decision maker's preferences. These methods can capture non-supported solutions, but they are usually computationally more expensive.

Quantum computing has emerged as a promising paradigm for solving certain classes of computational problems by exploiting quantum mechanical phenomena, such as superposition and entanglement \cite{Chuang2010}. Among quantum algorithms, variational quantum algorithms (VQAs) \cite{Cerezo2021} have received significant attention due to their compatibility with current noisy quantum devices. These algorithms combine a parameterized quantum circuit, executed on a quantum processor, with a classical optimization loop that updates the circuit parameters to minimize or maximize a given objective function. The Quantum Approximate Optimization Algorithm (QAOA) \cite{Farhi2014} is one of the most studied VQAs for combinatorial optimization problems, where it aims to find high-quality solutions by encoding the problem into a quantum circuit.

Recently, quantum approaches for MOO have started to emerge, particularly based on variational quantum algorithms such as QAOA. Previous quantum approaches typically relied on classical scalarization methods, such as weighted-sum or Tchebycheff formulations, where each preference vector is converted into a different single-objective optimization problem \cite{Chiew2024, Dahi2024}. Although conceptually straightforward, these methods require re-optimizing the QAOA parameters for every scalarized instance, creating a significant computational bottleneck. A recent work addressed this limitation by proposing a parameter-transfer strategy, in which QAOA parameters are optimized offline using quantum simulations on representative small instances and subsequently reused across multiple scalarizations of larger problems \cite{Kotil2025}. This approach reduces the need for repeated parameter optimization when solving different scalarized problems on quantum hardware. The authors demonstrated the effectiveness of this approach on multi-objective MAX-CUT instances using Matrix Product State (MPS) simulations and experiments on IBM quantum hardware. However, their study focused on MAX-CUT graphs matching the connectivity of the target quantum device, serving primarily as a proof of concept for the parameter-transfer strategy. Whether the same approach remains effective for realistic combinatorial optimization problems with more complex Hamiltonians is still an open question.

In this work, we investigate the applicability of this quantum MOO framework to multi-objective logistics problems, focusing on the Traveling Salesman Problem (TSP) and the Vehicle Routing Problem (VRP). We first formulate both problems as Quadratic Unconstrained Binary Optimization (QUBO) models and subsequently reduce them to equivalent MAX-CUT instances. The resulting graphs are then solved using the QAOA-based parameter-transfer framework proposed by \cite{Kotil2025}, which repeatedly samples weighted combinations of the objectives to approximate the Pareto front. This enables the generation of routing solutions representing different trade-offs among the considered objectives while avoiding the need to independently optimize QAOA parameters for each scalarized instance.

Following an a posteriori MOO strategy, our goal is to assess whether this approach can efficiently generate diverse high-quality routing solutions with different trade-offs among relevant logistics objectives, such as travel distance, delivery time, and route-related risk. By doing so, we evaluate the effectiveness of parameter transfer beyond proof-of-concept MAX-CUT instances, considering practical routing problems with substantially more structured and constrained optimization landscapes. To demonstrate the practical viability of this approach, we validate the parameter transfer strategy not only through comprehensive noise-free simulations but also via empirical experiments on real IBM quantum hardware.

Our experimental results show that parameter transfer remains effective in this more realistic setting, frequently producing Pareto fronts with a better coverage of the objective space than the adapted classical $\epsilon$-constraint method, while often reaching competitive solutions earlier under the computational assumptions adopted in our experiments. At the same time, we observe that its effectiveness strongly depends on the problem structure: the approach is considerably more stable for the TSP than for the more constrained VRP, suggesting that the concentration of transferable QAOA parameters decreases as the number of constraints and interactions in the underlying Hamiltonian increases.

This paper is organized as follows. Section \ref{sec:moo} presents the field of Multi-Objective Optimization, introducing concepts such as Pareto front, dominated and non-dominated solutions, supported and non-supported solutions, and hypervolume. Section \ref{sec:qamoo} describes the Quantum Approximate Multi-Objective Optimization approach and how we applied it to the TSP and VRP. Section \ref{sec:qubo} presents the QUBO formulations for the TSP and VRP, as well as the reduction to MAX-CUT problem. Section \ref{sec:results} presents the obtained results and, finally, Section \ref{sec:conclusion} concludes the paper and discusses future work.

\section{Multi-Objective Optimization}
\label{sec:moo}

In multi-objective optimization, the goal is to simultaneously optimize multiple objective functions $f_i$ over a feasible domain $X \subseteq \mathbb{R}^n$, that is,

\begin{equation}
    \max_{x \in X} \left[f_1(x), \ldots, f_m(x)\right],
\end{equation}

where $m \geq 2$ denotes the number of objective functions.

An optimal solution for this problem would be one that simultaneously maximizes all objective functions, that is,
\begin{equation}
    x^* = \arg\max_{x \in X} f_i(x), \quad \forall i \in \{1,\ldots,m\}.
\end{equation}
However, such a solution generally does not exist in MOO problems, since the objective functions are often conflicting, meaning that improving one objective may lead to the degradation of another. 

Therefore, instead of searching for a single globally optimal solution, MOO aims to identify a set of solutions that represent different trade-offs among the objectives. This is achieved through the concept of non-dominated solutions, where no objective can be improved without degrading at least one of the others. Given two feasible solutions $x, y \in X$, we say that $x$ dominates $y$ if

\begin{equation}
    f_i(x) \geq f_i(y), \quad \forall i \in \{1,\ldots,m\},
\end{equation}

and

\begin{equation}
    \exists j \in \{1,\ldots,m\} \text{ such that } f_j(x) > f_j(y).
\end{equation}

A solution $x^* \in X$ is said to be non-dominated if there exists no other feasible solution $x \in X$ that dominates it. Non-dominated solutions are also called Pareto-optimal solutions \cite{Hwang1979}.

\begin{figure}
    \centering
    \includegraphics[width=0.5\linewidth]{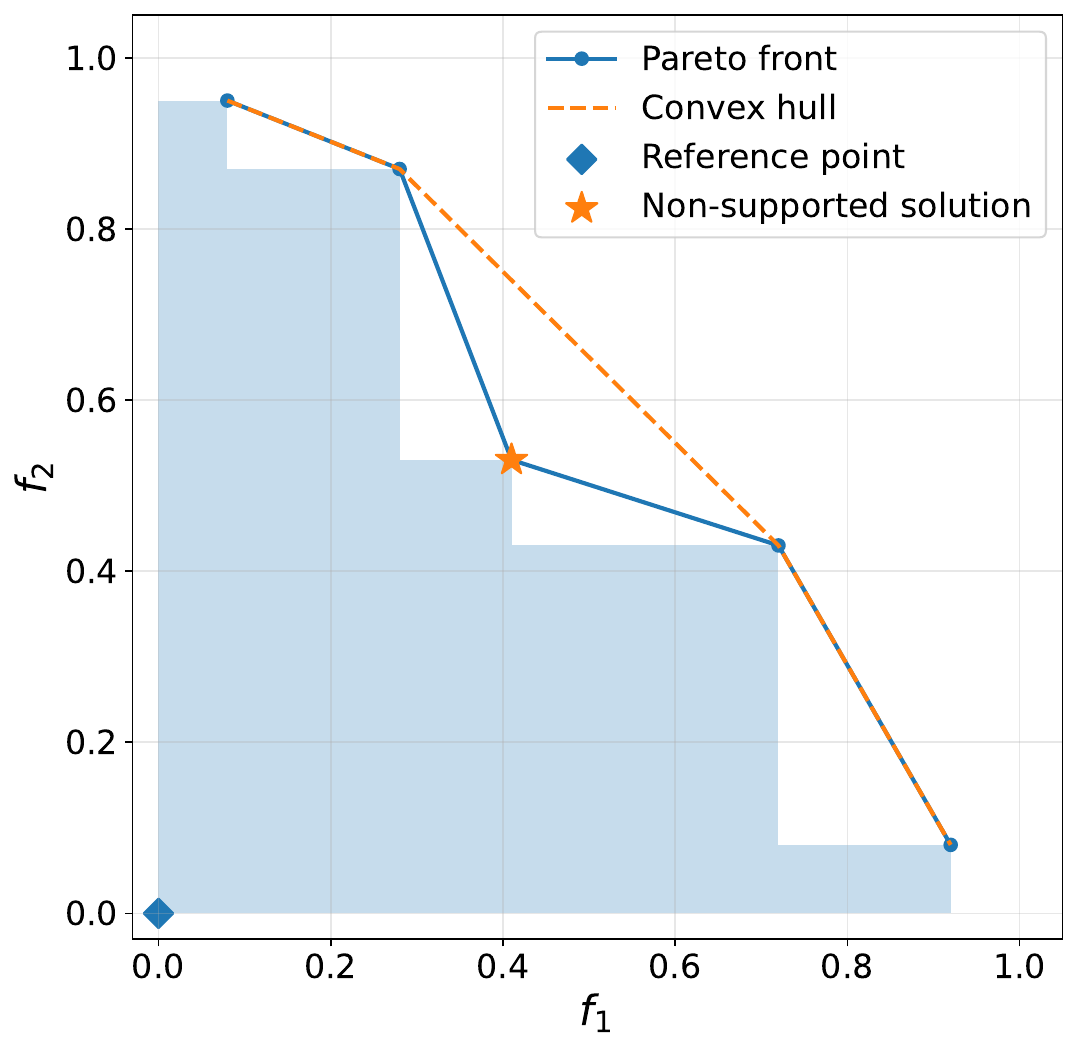}
    \caption{Illustrative two-objective Pareto front in a maximization problem. The shaded region represents the hypervolume dominated by the Pareto front with respect to the reference point. Solutions lying on the convex hull are supported and can be obtained by weighted-sum scalarization, whereas the highlighted interior point is Pareto-optimal but non-supported.}
    \label{fig:pareto_front}
\end{figure}

The set of all Pareto-optimal solutions in the decision space is called the Pareto-optimal set. Its image in the objective space is called the Pareto front. The Pareto front can be visualized as a set of points in an $m$-dimensional space, where each axis represents one objective function (see Figure \ref{fig:pareto_front}). Each point in the front corresponds to a Pareto-optimal solution and represents a different trade-off among the objectives. Points closer to the ideal point, i.e., the point that maximizes all objectives simultaneously, generally correspond to better trade-offs.

A common metric used to evaluate the quality of a Pareto front approximation is the hypervolume (HV). Given a reference point in the objective space, the hypervolume measures the volume of the region dominated by the Pareto front and bounded by this reference point. A larger hypervolume usually indicates a better approximation of the Pareto front, as it reflects both good convergence toward the optimal front and a diverse spread of solutions across the objective space \cite{Ehrgott2005}. In Figure~\ref{fig:pareto_front}, the hypervolume corresponds to the shaded area dominated by the Pareto front with respect to the reference point.

In the Pareto front, solutions can also be classified as supported and non-supported solutions. A Pareto-optimal solution is said to be supported if it can be obtained as the optimal solution of the \emph{Weighted Sum Method} (WSM), in which the multi-objective problem is scalarized through a convex combination of the objective functions. More formally, there exists a weight vector $c = (c_1, \ldots, c_m)$ with $c_i > 0$ such that

\begin{equation}
    x^* = \arg\max_{x \in X} \sum_{i=1}^{m} c_i f_i(x).
\end{equation}

Geometrically, supported solutions lie on the convex boundary of the Pareto front and can be found by moving a supporting hyperplane over the objective space. On the other hand, non-supported solutions are Pareto-optimal solutions that cannot be obtained by the WSM. These solutions typically lie in non-convex regions of the Pareto front and represent trade-offs that are missed by classical scalarization-based approaches \cite{Ehrgott2005}.

\section{QUBO Formulations}
\label{sec:qubo}

Many combinatorial optimization problems can be reformulated as Quadratic Unconstrained Binary Optimization (QUBO) problems, which constitute a standard representation for quantum optimization algorithms. A QUBO problem is defined as the optimization of a quadratic function over binary variables:

\begin{equation}
    \max_{x \in \{0,1\}^n} x^T Q x,
\end{equation}

where $x$ is a binary vector and $Q \in \mathbb{R}^{n \times n}$ is a matrix encoding both the objective function and the problem constraints through penalty terms \cite{Glover2022}.

Because QUBO models can be directly mapped to Ising Hamiltonians, they constitute the starting point for many variational quantum optimization methods, including the framework adopted in this work. Our approach first formulates the TSP and VRP as QUBO problems, then transforms these formulations into equivalent MAX-CUT instances, which are subsequently solved using a QAOA-based parameter-transfer strategy to approximate the Pareto front.

\subsection{QUBO formulation of TSP}

Let $C={0,\ldots,n-1}$ denote the set of custumers and $T={0,\ldots,n-1}$ the set of positions in the tour. We introduce a binary decision variable

\begin{equation}
x_{i,t}=
\begin{cases}
1, & \text{if city } i \text{ is visited at position } t,\\
0, & \text{otherwise},
\end{cases}
\end{equation}

for every $i\in C$ and $t\in T$. Therefore, the formulation contains $n^2$ binary variables.

The objective is to minimize the total travel cost of the tour. Let $d_{ij}$ denote the travel cost from city $i$ to city $j$. Since consecutive positions in the tour correspond to consecutive visited cities, the travel cost can be written as

\begin{equation}\label{eq:tsp-obj}
\sum_{t=0}^{n-1}\sum_{i\in C}\sum_{\substack{j\in C \, j\neq i}}
d_{ij}\,x_{i,t}\,x_{j,(t+1)\bmod n},
\end{equation}

where the modulo operator ensures that the last city is connected back to the first one, forming a Hamiltonian cycle.

To guarantee that the binary variables represent a valid tour, two sets of constraints are required. First, each position in the tour must be occupied by exactly one city:

\begin{equation}
\sum_{i\in C} x_{i,t}=1,
\qquad \forall t\in T.
\end{equation}

Second, each city must appear exactly once in the tour:

\begin{equation}
\sum_{t\in T} x_{i,t}=1,
\qquad \forall i\in C.
\end{equation}

Since QUBO problems are unconstrained, these constraints are incorporated into the objective function through quadratic penalty terms. The resulting QUBO formulation is given by

\begin{equation}
\min_x
\;
\sum_{t=0}^{n-1}\sum_{i\in C}\sum_{\substack{j\in C \, j\neq i}}
d_{ij}\,x_{i,t}\,x_{j,(t+1)\bmod n}
+
\lambda_1
\sum_{t\in T}
\left(
\sum_{i\in C}x_{i,t}-1
\right)^2
+
\lambda_2
\sum_{i\in C}
\left(
\sum_{t\in T}x_{i,t}-1
\right)^2,
\end{equation}

where $\lambda_1$ and $\lambda_2$ are penalty coefficients chosen sufficiently large so that infeasible solutions become energetically unfavorable compared to feasible tours \cite{Lucas2014}[Section 7]. In our implementation, these penalty coefficients are scaled according to the maximum edge cost of the instance, balancing the contribution of the objective function and the constraint penalties.

\subsection{QUBO formulation of VRP}

The Vehicle Routing Problem (VRP) extends the TSP by considering multiple vehicles departing from a common depot to serve a set of customers. Our formulation follows the same time-indexed modeling strategy adopted for the TSP, extending it with an additional vehicle index inspired by classical time-indexed formulations for the VRP \cite{vrp2021}.

Let $K={1,\ldots,m}$ denote the set of vehicles, $C={1,\ldots,n}$ the set of customers, and let node $0$ represent the depot. Unlike the TSP, we introduce an auxiliary node $\emptyset$, representing the end of a route. Therefore, the formulation considers a total of $n+2$ nodes: the depot, the $n$ customers, and the auxiliary node.

We define the binary decision variables

\begin{equation}
x_{i,k,t}=
\begin{cases}
1, & \text{if vehicle } k \text{ visits node } i \text{ at position } t,\\
0, & \text{otherwise},
\end{cases}
\end{equation}

for every vehicle $k\in K$, node $i$, and route position $t$. Consequently, the formulation contains $m(n+2)T$ binary variables, where $T$ denotes the maximum number of positions allowed in each route.

The auxiliary node allows routes to terminate before reaching the maximum route length while preserving a fixed number of decision variables. Once a vehicle reaches node $\emptyset$, all subsequent positions in its route must also be assigned to $\emptyset$, indicating that the route has ended. Unlike the TSP, vehicles are not required to return to the depot. 

The travel-cost objective is analogous to the TSP formulation, minimizing the total routing cost over all vehicles. Let ($c_{ij}$) denote the travel cost from node ($i$) to node ($j$). The objective function is given by

\begin{equation}
\sum_{k\in K}\sum_{t=0}^{T-2}
\sum_{i\in V}\sum_{j\in V}
c_{ij} \, x_{i,k,t} \, x_{j,k,t+1},
\end{equation}

where ($V=\{0\}\cup C\cup\{\emptyset\}$) is the set of all nodes. Transitions from a customer to the auxiliary node ($\emptyset$) are assigned the cost of returning from that customer to the depot, whereas transitions from ($\emptyset$) to itself have zero cost.

As in the TSP formulation, the constraints are incorporated into the objective function through quadratic penalty terms. The first two constraints are direct extensions of the TSP formulation. First, each vehicle must occupy exactly one node at every route position,

\begin{equation}
\sum_{i\in V} x_{i,k,t}=1,
\qquad
\forall k\in K, \; t=0,\ldots,T-1,
\end{equation}

where ($V=\{0\}\cup C\cup\{\emptyset\}$) denotes the complete set of nodes.

Second, every customer must be visited exactly once by exactly one vehicle,

\begin{equation}
\sum_{k\in K}\sum_{t=0}^{T-1}x_{i,k,t}=1,
\qquad
\forall i\in C.
\end{equation}

Unlike the TSP, every vehicle must start its route at the depot. This is enforced by

\begin{equation}
x_{0,k,0}=1,
\qquad
\forall k\in K.
\end{equation}

Since vehicles leave the depot only once, the depot cannot appear again in the remaining positions of the route,

\begin{equation}
x_{0,k,t}=0,
\qquad
\forall k\in K, \; t>0.
\end{equation}

To ensure that the auxiliary node correctly represents the end of a route, once a vehicle reaches node ($\emptyset$), all subsequent positions must also correspond to ($\emptyset$). This continuity condition is enforced through the penalty

\begin{equation}
x_{\emptyset,k,t}-x_{\emptyset,k,t+1}\leq 0,
\qquad
\forall k\in K, \; t=0,\ldots,T-2,
\end{equation}

which guarantees that the sequence of auxiliary nodes is contiguous until the end of the route.

Finally, every vehicle is required to serve at least one customer. Since the depot occupies the first position, this condition can be written as

\begin{equation}
\sum_{i\in C}\sum_{t=1}^{T-1}x_{i,k,t}\geq 1,
\qquad
\forall k\in K.
\end{equation}

Each of the above constraints is incorporated into the objective function through quadratic penalty terms, yielding the following QUBO formulation:

\begin{align}
Q_{\mathrm{VRP}} = & \sum_{k\in K}\sum_{t=0}^{T-2}\sum_{i\in V}\sum_{j\in V} 
c_{ij}\,x_{i,k,t}\,x_{j,k,t+1} \nonumber \\
& + \lambda_1 \sum_{k\in K}\sum_{t=0}^{T-1}
\left(\sum_{i\in V}x_{i,k,t}-1\right)^2 \nonumber \\
& + \lambda_2 \sum_{i\in C}
\left(\sum_{k\in K}\sum_{t=0}^{T-1}x_{i,k,t}-1\right)^2 \nonumber \\
& + \lambda_3 \sum_{k\in K}(x_{0,k,0}-1)^2 \nonumber \\
& + \lambda_4 \sum_{k\in K}\sum_{t=1}^{T-1}
x_{0,k,t} \nonumber \\
& + \lambda_5 \sum_{k\in K}\sum_{t=0}^{T-2}
\left(x_{\emptyset,k,t} - x_{\emptyset,k,t}x_{\emptyset,k,t+1}\right) \nonumber \\
& + \lambda_6 \sum_{k\in K} \left(\sum_{i\in C}\sum_{t=1}^{T-1}x_{i,k,t}-1\right)^2.
\end{align}

where $\lambda_1,\ldots,\lambda_6$ are penalty coefficients chosen sufficiently large so that infeasible solutions become energetically unfavorable compared to feasible routes. As in the TSP formulation, these coefficients are scaled according to the maximum edge cost of the corresponding instance, balancing the contribution of the objective function and the constraint penalties.

\subsection{Reduction from QUBO to MAX-CUT}

The QAOA framework adopted in this work operates on MAX-CUT instances. Therefore, the QUBO formulations of the TSP and VRP are transformed into equivalent weighted MAX-CUT problems following the reduction proposed by \cite{Bhattacharyya2025}.

Given a QUBO matrix $Q\in\mathbb{R}^{n\times n}$, we first transform it into its equivalent symmetric representation,
\begin{equation}
Q
\leftarrow
\frac{Q+Q^T}{2},
\end{equation}
because the antisymmetric component of $Q$ does not contribute to the quadratic form. Therefore, the original QUBO objective remains unchanged, while the resulting matrix is symmetric and suitable for the QUBO-to-MAX-CUT reduction.

Next, the matrix is normalized according to

\begin{equation}
Q
\leftarrow
\frac{Q}
{\displaystyle\max_{i,j}|Q_{ij}|},
\end{equation}

so that all coefficients lie in the interval $[-1,1]$. This normalization improves the numerical conditioning of the resulting MAX-CUT instance and ensures that all optimization instances are represented on a comparable scale.

The normalized QUBO matrix is then converted into a weighted graph
$G=(V,E)$.
The graph contains $n+1$ vertices, where the first $n$ vertices correspond to the binary decision variables and the additional vertex is introduced by the reduction.

The weighted adjacency matrix $A$ is defined as

\begin{equation}
A_{ij}
=
-Q_{ij},
\qquad
i\neq j,
\end{equation}

for all pairs of decision-variable vertices, while the edges connecting each variable vertex to the auxiliary vertex are assigned weights

\begin{equation}
A_{i,n+1}
=
\sum_{j=1}^{n}Q_{ij}.
\end{equation}

The resulting graph is fully connected, since every pair of binary variables may interact through a quadratic coefficient of the QUBO formulation. Solving the corresponding weighted MAX-CUT instance is therefore equivalent to optimizing the original QUBO problem, allowing the QAOA methodology described in the previous sections to be directly applied.

\section{Quantum Approximate Multi-Objective Optimization}
\label{sec:qamoo}

To solve the multi-objective routing problems considered in this work, we adopt the QAOA-based framework proposed by \cite{Kotil2025}. Here, we consider multi-objective TSP and VRP instances in which the objectives differ only in their cost matrices, representing travel distance, travel time, and risk. Since all objectives share the same decision variables and constraints, the resulting QUBO formulations differ only in the coefficients of the objective function, while their structure remains unchanged. These QUBO formulations are then reduced to equivalent MAX-CUT instances, allowing the application of the QAOA-based optimization framework. The method approximates the Pareto front by repeatedly solving scalarized MAX-CUT instances obtained from randomly sampled convex combinations of the objectives. To make this procedure computationally efficient, the variational parameters of QAOA are optimized only once on a representative training instance and subsequently transferred to all scalarized target instances, eliminating the need for repeated classical optimization.

The following subsections briefly review the QAOA algorithm and then describe the multi-objective optimization framework adopted in this work.

\subsection{Quantum Approximate Opimization Algorithm}

The Quantum Approximate Optimization Algorithm (QAOA) \cite{Farhi2014} is a hybrid quantum-classical variational algorithm designed to solve combinatorial optimization problems. The algorithm relies on two complementary Hamiltonians: a cost Hamiltonian $H_C$, which encodes the optimization problem, and a mixer Hamiltonian $H_M$, which explores the search space by driving transitions between candidate solutions. Starting from an initial quantum state that equally represents all feasible binary assignments, QAOA alternates the application of these two Hamiltonians for a fixed number of layers, known as the circuit depth $p$. Increasing the depth allows the algorithm to explore more expressive quantum states, potentially improving solution quality, but also increases the number of variational parameters and the circuit complexity. The resulting quantum state is parameterized by a set of variational angles, which are optimized by a classical optimizer to minimize the expected value of the cost Hamiltonian. After the optimization converges, measurements of the final quantum state produce candidate solutions, with lower-energy (higher-quality) solutions being sampled with higher probability.

The optimization problems considered in this work are formulated as QUBO models (Section~\ref{sec:qubo}). The corresponding QUBO objective is mapped to an Ising Hamiltonian through the transformation

\begin{equation}
    x_i = \frac{1-z_i}{2},
\end{equation}

where $z_i \in \{-1,1\}$. This mapping converts the QUBO formulation into an equivalent Ising Hamiltonian, which serves as the cost Hamiltonian of the QAOA,

\begin{equation}
    H_C = - \sum_{i<j} J_{ij}\sigma_i^z\sigma_j^z
          - \sum_i h_i\sigma_i^z,
\end{equation}

where $\sigma_i^z$ denotes the Pauli-$Z$ operator acting on qubit $i$, while $J_{ij}$ and $h_i$ are coefficients derived from the QUBO matrix. Consequently, minimizing the energy of $H_C$ is equivalent to maximizing the original QUBO objective function.

The algorithm starts from the uniform superposition state,

\begin{equation}
    |+\rangle =
    \frac{1}{\sqrt{2^n}}
    \sum_{x\in\{0,1\}^n}|x\rangle,
\end{equation}

which corresponds to the ground state of the mixer Hamiltonian

\begin{equation}
    H_M = -\sum_i \sigma_i^x,
\end{equation}

where $\sigma_i^x$ denotes the Pauli-$X$ operator. QAOA alternates the application of the cost and mixer Hamiltonians for $p$ layers, producing the variational quantum state

\begin{equation}
    |\psi(\boldsymbol{\beta},\boldsymbol{\gamma})\rangle =
    \prod_{l=1}^{p}
    e^{-i\beta_l H_M}
    e^{-i\gamma_l H_C}
    |+\rangle,
\end{equation}

where $\boldsymbol{\beta}=(\beta_1,\ldots,\beta_p)$ and $\boldsymbol{\gamma}=(\gamma_1,\ldots,\gamma_p)$ are variational parameters. The objective is to determine the parameter values that minimize the expectation value of the cost Hamiltonian,

\begin{equation}
    \min_{\boldsymbol{\beta},\boldsymbol{\gamma}}
    \left\langle
    \psi(\boldsymbol{\beta},\boldsymbol{\gamma})
    \right|
    H_C
    \left|
    \psi(\boldsymbol{\beta},\boldsymbol{\gamma})
    \right\rangle.
\end{equation}

The optimization is performed using a classical optimizer, while the quantum computer prepares and measures the variational state. After convergence, repeated measurements of the optimized state produce a probability distribution over candidate solutions, where low-energy states—and therefore high-quality solutions to the original optimization problem—are sampled with higher probability.

\subsection{Multi-objective optimization with QAOA}

Once the multi-objective routing problem has been reduced to a multi-objective MAX-CUT instance, it is solved using the framework proposed by \cite{Kotil2025}. The method repeatedly applies WSM to generate scalarized optimization problems, each corresponding to a different trade-off among the objectives.

Let $f_1,\ldots,f_m$ denote the objective functions of the multi-objective MAX-CUT problem. For each iteration, a weight vector $\mathbf{c}=(c_1,\ldots,c_m)$ is sampled from the unit simplex:

\begin{equation}
c_i\ge0,
\qquad
\sum_{i=1}^{m}c_i=1.
\end{equation}

The sampled vector defines the scalarized objective

\begin{equation}
f_{\mathbf{c}}(x)
=
\sum_{i=1}^{m}c_i f_i(x),
\end{equation}

which is converted into the corresponding cost Hamiltonian. QAOA then optimizes its variational parameters to minimize the expectation value of this Hamiltonian, producing a candidate solution for the scalarized problem. Repeating this procedure for multiple weight vectors produces solutions corresponding to different trade-offs among the objectives, allowing the Pareto front to be approximated \cite{Ehrgott2005}.

When the scalarized problems are solved exactly using a classical optimization algorithm, the resulting solutions are guaranteed to be Pareto-optimal. However, the WSM can only recover supported Pareto solutions \cite{Ehrgott2005}. The proposed approach overcomes this limitation by exploiting the probabilistic nature of QAOA. Instead of returning only the optimal solution of each scalarized problem, QAOA prepares a quantum state whose measurement produces a probability distribution over multiple candidate solutions. Consequently, a single scalarized problem may generate several feasible solutions, including high-quality solutions that are not optimal for the corresponding weighted objective, and some of these solutions may correspond to non-supported Pareto-optimal solutions. Therefore, increasing the number of sampled weight vectors explores a wider range of trade-offs among the objectives, while increasing the number of circuit measurements (shots) increases the likelihood of sampling non-supported Pareto-optimal solutions.

After evaluating all sampled solutions with respect to the original objective functions, the dominance relation is applied to identify the non-dominated solutions. The resulting set forms an approximation of the Pareto front, which is then presented to the decision maker. 

\subsection{Parameter transfer}

Recent studies have shown that the variational parameters of QAOA exhibit a degree of transferability across related optimization instances \cite{Brandao2018,Shaydulin2023,Galda2023,Katial2025}. This observation is particularly relevant because the computational cost of QAOA is largely dominated by the repeated classical optimization of these parameters. Brandão et al. \cite{Brandao2018} demonstrated that, for typical problem distributions, the QAOA objective landscapes associated with different problem instances become increasingly similar, allowing a single set of variational parameters to achieve good performance across multiple instances. Subsequent works extended this idea to weighted combinatorial optimization problems and proposed parameter-transfer strategies that substantially reduce the classical optimization effort while preserving solution quality \cite{Shaydulin2023,Galda2023,Katial2025}.

Following this line of work, we adopt the parameter-transfer strategy proposed by \cite{Kotil2025}. Instead of optimizing the QAOA parameters for every scalarized MAX-CUT instance, the parameters are trained once on a representative MAX-CUT instance and subsequently reused for all scalarized MAX-CUT problems.

Let $(\boldsymbol{\beta},\boldsymbol{\gamma})=(\beta_1,\ldots,\beta_p,\gamma_1,\ldots,\gamma_p)$ denote the QAOA parameters. These parameters are first optimized offline on a representative MAX-CUT training graph that is sufficiently small to be simulated classically. This graph has the same topology as the target MAX-CUT graph obtained from the QUBO reduction, while its edge weights are independently sampled according to

\begin{equation}
w_{ij}\sim\mathcal N\!\left(0,\frac{1}{m}\right),
\end{equation}

where \(m\) is the number of objective functions. As shown by \cite{Kotil2025}, this choice produces transferable QAOA parameters that can be reused for the scalarized target instances. 

The parameters are obtained by solving

\begin{equation}
(\boldsymbol{\beta}^{*},\boldsymbol{\gamma}^{*})
=
\arg\min_{\boldsymbol{\beta},\boldsymbol{\gamma}}
\left\langle
\psi(\boldsymbol{\beta},\boldsymbol{\gamma})
\right|
H_{\mathrm{train}}
\left|
\psi(\boldsymbol{\beta},\boldsymbol{\gamma})
\right\rangle,
\end{equation}

where $H_{\mathrm{train}}$ denotes the Hamiltonian associated with the training instance.

Once the parameters have been determined, they remain fixed throughout the optimization process. For each randomly sampled weight vector $\mathbf{c}$, only the cost Hamiltonian changes,

\begin{equation}
H(\mathbf{c})
=
\sum_{i=1}^{m}
c_i H_i,
\end{equation}

where $H_i$ is the Hamiltonian corresponding to the $i$-th objective function. The QAOA circuit is then executed using the transferred parameters $(\boldsymbol{\beta}^{*},\boldsymbol{\gamma}^{*})$, without performing any additional classical optimization.

This strategy decouples the parameter optimization from the solution of the target problem. Since the parameters are computed offline on a representative instance, the online phase consists solely of evaluating different scalarizations by sampling weight vectors $\mathbf{c}$ and executing the corresponding QAOA circuits.

\section{Results and Discussion}
\label{sec:results}

This section evaluates the proposed methodology on multi-objective instances of the TSP and the VRP. TSP instances containing 4, 5, and 6 customers were considered, together with VRP instances containing 4 and 5 customers served by 1 vehicle and 3 customers served by 1 and 2 vehicles. For each instance size, three independent random seeds were utilized. The selected problem sizes correspond to the largest instances that could be simulated with the available computational resources. The corresponding QUBO formulations were converted into MAX-CUT instances as described in Section~\ref{sec:qubo}. The evaluated instances and their corresponding number of qubits are summarized in Table~\ref{tab:instances}. As described in Section~\ref{sec:qubo}, the number of qubits required by QAOA is determined by the number of binary variables in the QUBO formulation. Since the QUBO-to-MAX-CUT reduction introduces one additional variable, the total number of qubits is $|C|^2+1$ for the TSP and $|K|(|C|+2)T+1$ for the VRP.

\begin{table}[ht]
\centering
\caption{Problem instances considered in the computational experiments.}
\label{tab:instances}
\begin{tabular}{cccc}
\hline
Problem & Customers (C) & Vehicles (K) & Qubits \\
\hline
\multirow{3}{*}{TSP}
& 4 & -- & 17 \\
& 5 & -- & 26 \\
& 6 & -- & 37 \\
\hline
\multirow{4}{*}{VRP}
& 3 & 1 & 21 \\
& 3 & 2 & 41 \\
& 4 & 1 & 31 \\
& 5 & 1 & 43 \\
\hline
\end{tabular}
\end{table}

Each routing instance was defined by three objective functions representing travel distance, travel time, and transportation risk. Customer coordinates were uniformly sampled in a two-dimensional plane, and the distance objective was computed from the Euclidean distances between locations. Travel times were obtained by multiplying the Euclidean distances by randomly generated congestion factors, while the risk objective was generated as a symmetric matrix of integer values representing the relative risk associated with each edge. Since the objectives differ only in their cost matrices, all instances share the same decision variables and constraints.

Following the parameter-transfer strategy proposed by \cite{Kotil2025}, QAOA parameters were trained offline on representative random MAX-CUT instances. The training graphs were generated as fully connected weighted graphs whose edge weights were independently sampled according to the distribution $\mathcal{N}(0,1/3)$, corresponding to the three-objective setting considered in this work. To preserve the training-to-target scaling adopted by \cite{Kotil2025}, where parameters trained on a 27-qubit instance were transferred to a 42-qubit problem, we selected training graph sizes that approximately maintain the same qubit ratio with respect to the target routing instances. The classical parameter optimization was performed using the JuliQAOA package \cite{Golden2023JuliQAOA}, following the same optimization procedure adopted in \cite{Kotil2025}. The evaluated parameter-transfer configurations are summarized in Table~\ref{tab:transfer}.

\begin{table}[ht]
\centering
\caption{Training and target graph sizes used in the parameter-transfer experiments.}
\label{tab:transfer}
\begin{tabular}{lcc}
\hline
Instance & Target qubits & Training qubits \\
\hline
TSP-4   & 17 & 7, 11 \\
TSP-5   & 26 & 11, 17 \\
TSP-6   & 37 & 17, 27 \\
\hline
VRP-3-1 & 21 & 8, 14 \\
VRP-4-1 & 31 & 14, 20 \\
VRP-3-2 & 41 & 20, 27 \\
VRP-5-1 & 43 & 20, 27 \\
\hline
\end{tabular}
\end{table}

All experiments were executed on an Amazon Web Services (AWS) EC2 \texttt{c8i.8xlarge} instance equipped with an Intel(R) Xeon(R) 6975P-C processor (32 vCPUs) and 64 GB of RAM. Quantum simulations employed the Matrix Product State (MPS) simulator available in Qiskit v1.3.2 with bond dimension $\chi=20$, which was shown by \cite{Kotil2025} to provide sufficient accuracy for the considered problem sizes. The QAOA depth was varied from $p=1$ to $p=6$. For each experiment, 5,000 weight vectors were uniformly sampled from the unit simplex, and the optimized quantum circuit was measured 5,000 times for each scalarized instance. For comparison, the adapted $\epsilon$-constraint method proposed by \cite{Kotil2025} was also executed using the same computational budget adopted for the QAOA experiments.

\begin{figure}[t]
    \centering
    \includegraphics[width=1\linewidth]{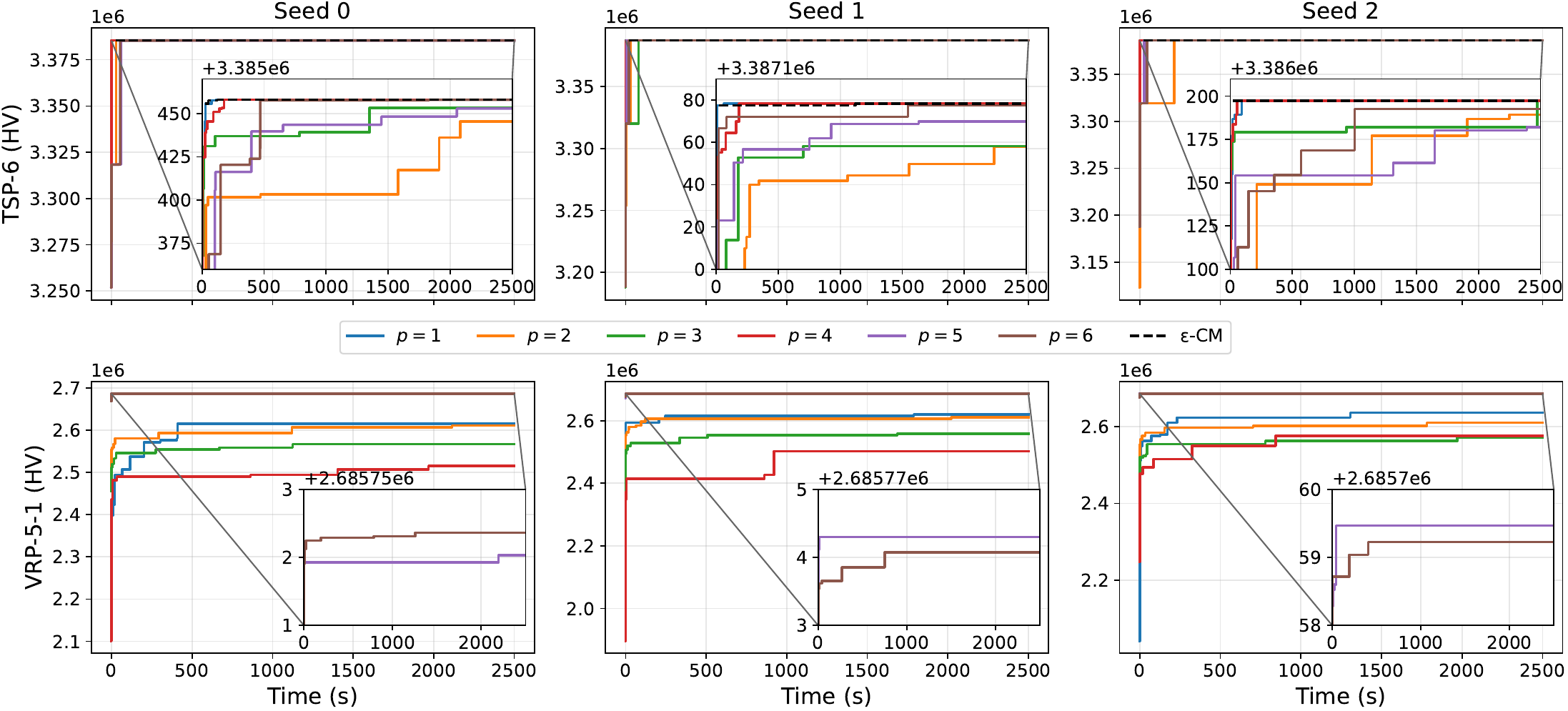}
    \caption{Hypervolume evolution over time for different QAOA depths ($p$) on the TSP-6 and VRP-5-1 instances using parameters transferred from 27-qubit training graphs. Each column corresponds to a different random seed. The vertical axis limits are independently adjusted for each subplot to improve visualization of the small variations in hypervolume among different configurations. For the TSP, the dashed black line denotes the performance of the adapted $\epsilon$-constraint method under the same 2,500 s computational budget. The $\epsilon$-constraint method is not reported for the VRP instance because the problem size exceeds the limits imposed by the available Gurobi license. Hypervolume values are reported in arbitrary units (A.U.).}
    \label{fig:hv_progress}
\end{figure}

Figure~\ref{fig:hv_progress} illustrates the HV evolution throughout the optimization process for the largest TSP and VRP instances considered in this work, namely TSP-6 and VRP-5-1. These instances represent the most challenging cases evaluated in terms of problem size and QUBO dimensionality. The horizontal axis represents the execution time assuming a sampling rate of 10,000 measurement shots per second. Since each experiment evaluates 5,000 scalarization weights using 5,000 measurement shots each, the total computational budget corresponds to 25 million shots, or 2,500 seconds. Overall, the experiments show that parameter transfer enables the generation of competitive Pareto approximations without requiring independent QAOA parameter optimization for each scalarized instance. However, the observed performance depends on the problem structure: TSP instances exhibit more consistent improvements across different QAOA depths, while VRP instances show larger variations due to the additional constraints introduced by the formulation. Equivalent plots for all remaining experiments are provided in the accompanying code repository.

Several observations can be made from these results. First, parameter transfer consistently enables the algorithm to reach the maximum hypervolume for the considered instances. Second, for the TSP-6 instance, depths $p=1$ and $p=4$ attain the maximum hypervolume, with $p=1$ reaching it first in all three random seeds. Moreover, all QAOA depths produce feasible TSP solutions. In contrast, the more constrained VRP-5-1 instance requires deeper circuits, with only $p=5$ and $p=6$ achieving the maximum hypervolume. The performance of these two depths is very similar across all three random seeds. However, only $p=6$ attains the maximum hypervolume for Seed 0, whereas only $p=5$ does so for Seeds 1 and 2. Furthermore, only these two depths produce feasible VRP solutions, while shallower circuits violate the problem constraints.

\begin{figure}[h]
    \centering
    \includegraphics[width=1\linewidth]{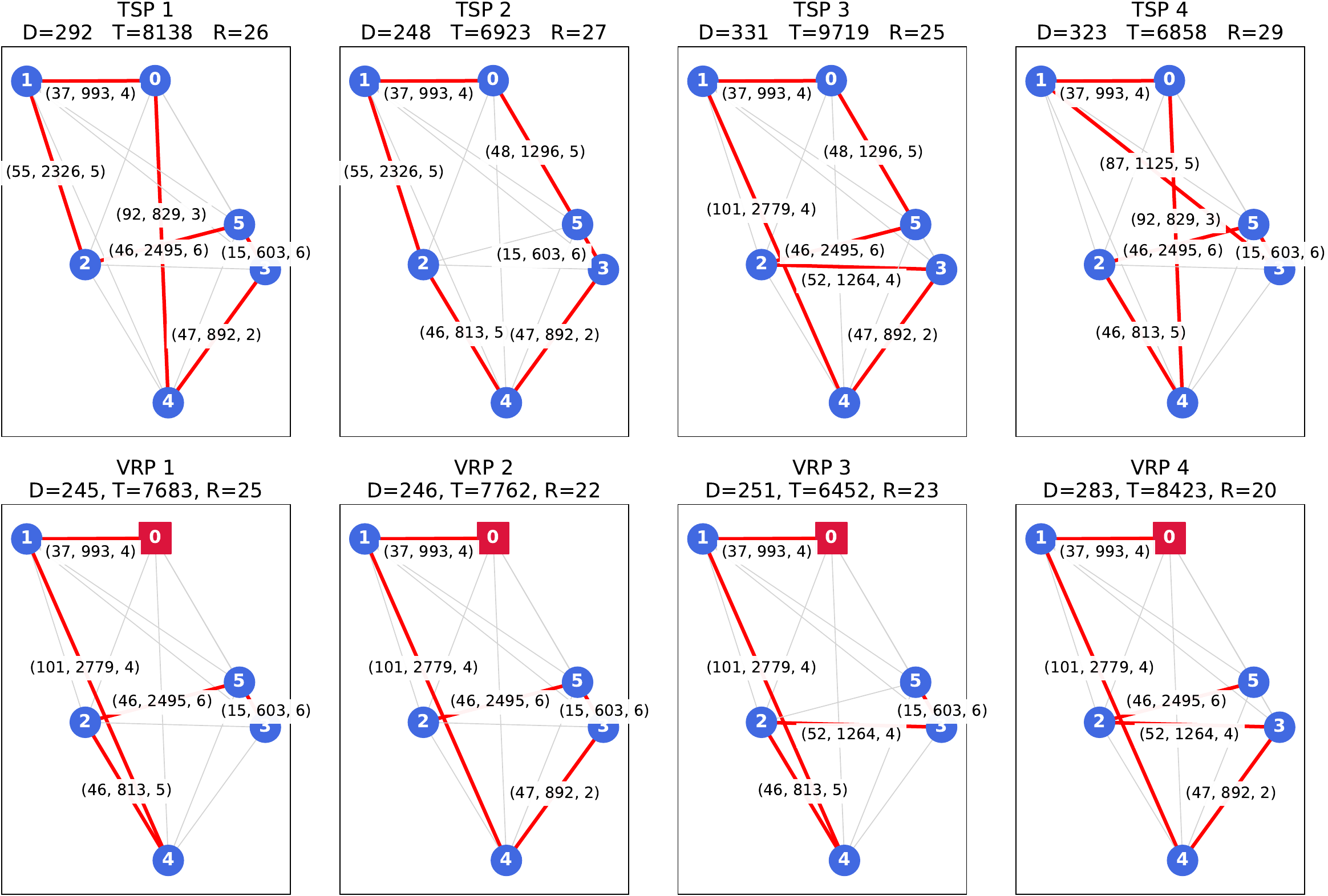}
    \caption{Representative non-dominated solutions obtained using parameter transfer from a 27-qubit training graph for the TSP-6 ($p=1$) and VRP-5-1 ($p=6$) instances. Each row corresponds to one problem and each column to a different solution from the approximated Pareto front. The objective values are shown above each solution, where D, T, and R denote distance, travel time, and risk, respectively. The TSP edge labels indicate the corresponding objective values of the selected route.}
    \label{fig:example-solutions}
\end{figure}

For the three random seeds, the obtained numbers of non-dominated solutions were 16, 4, and 5 for the TSP instance, and 5, 4, and 4 for the VRP instance, respectively. Figure~\ref{fig:example-solutions} illustrates the non-dominated solutions for the TSP-6 and VRP-5-1 instances with Seed 1. The objective values exhibit clear trade-offs, with improvements in one objective generally accompanied by deteriorations in at least one of the others. For the VRP, node 0 denotes the depot (shown as a square). Since the considered formulation corresponds to an open VRP, vehicles are not required to return to the depot after serving their assigned customers.

\begin{table*}[!h]
\centering
\caption{QAOA depths ($p$) yielding the highest hypervolume (Max HV) observed among all evaluated methods, QAOA depths, and training graph sizes for each problem instance and random seed. The second group reports the depths that generated feasible solutions. The last three columns report the number of non-dominated (ND) solutions obtained for each seed. Entries marked with an asterisk (*) indicate configurations whose highest hypervolume did not reach the maximum value obtained across all evaluated QAOA depths, training graph sizes, and comparison methods for the corresponding instance and seed.}
\label{tab:bestp}
\begin{tabular}{ll|ccc|ccc|ccc}
\hline
& &
\multicolumn{3}{c|}{Max HV $p$} &
\multicolumn{3}{c|}{Feasible $p$} &
\multicolumn{3}{c}{ND Solutions}\\
\cline{3-11}
Instance & Train &
S0 & S1 & S2 &
S0 & S1 & S2 &
S0 & S1 & S2\\
\hline

\multirow{2}{*}{TSP-4}
& 7  & All & All & All & All & All & All & \multirow{2}{*}{1} & \multirow{2}{*}{3} & \multirow{2}{*}{2}\\
& 11 & All & All & All & All & All & All & & &\\
\hline

\multirow{2}{*}{TSP-5}
& 11 & 1 & 1 & 1,6 & All & All & All & \multirow{2}{*}{4} & \multirow{2}{*}{9} & \multirow{2}{*}{3}\\
& 17 & 3--5 & 1,4,5 & 1,4,5 & All & All & All & & &\\
\hline

\multirow{2}{*}{TSP-6}
& 17 & 3,4 & 3 & 3,4 & 3--5 & 1--5 & 3--5 & \multirow{2}{*}{16} & \multirow{2}{*}{4} & \multirow{2}{*}{5}\\
& 27 & 1,4 & 1,4 & 1,4 & All & All & All & & &\\
\hline

\multirow{2}{*}{VRP-3-1}
& 8  & 3 & 1,2 & 1,2 & 1--3,6 & 1--3,5 & 1--3,5,6 & \multirow{2}{*}{3} & \multirow{2}{*}{5} & \multirow{2}{*}{4}\\
& 14 & 1--3 & 1,2,6 & 6 & 1--3,5 & All & 1--3,5,6 & & &\\
\hline

\multirow{2}{*}{VRP-4-1}
& 14 & 2* & 1* & 3* & 2 & 1 & 1,3,6 & \multirow{2}{*}{7} & \multirow{2}{*}{7} & \multirow{2}{*}{6}\\
& 20 & 1* & 1* & 1* & 1,6 & 1,6 & 1,2,5 & & &\\
\hline

\multirow{2}{*}{VRP-3-2}
& 20 & 2* & 3* & 2* & 1,4,6 & 1--4,6 & 2--4,6 & \multirow{2}{*}{2} & \multirow{2}{*}{3} & \multirow{2}{*}{2}\\
& 27 & 4--6 & 4 & 4--6 & 1,4--6 & 1,4--6 & All & & &\\
\hline

\multirow{2}{*}{VRP-5-1}
& 20 & 4* & 4* & 4* & -- & 4 & -- & \multirow{2}{*}{5} & \multirow{2}{*}{4} & \multirow{2}{*}{4}\\
& 27 & 6 & 5 & 5 & 5,6 & 5,6 & 5,6 & & &\\
\hline

\end{tabular}
\end{table*}

With respect to the convergence behavior, a difference can be observed between the two problem classes. While the TSP instances exhibit a rapid convergence toward the maximum hypervolume, the VRP instances require substantially longer execution times and present larger variability across QAOA depths. This behavior can be attributed to the additional variables and feasibility constraints introduced by the VRP formulation, which result in a more structured optimization landscape and a more constrained cost Hamiltonian.

This behavior is also observed in the remaining instances. Table~\ref{tab:bestp} summarizes the complete set of results. For each problem instance and random seed, the maximum hypervolume value was determined by comparing all evaluated QAOA depths, training graph sizes, and, when available, the adapted $\epsilon$-constraint method. The table reports the QAOA depths that achieved this maximum hypervolume, the depths that generated feasible solutions, and the number of non-dominated (ND) solutions obtained for each seed. 

For the smallest instance (TSP-4), every tested QAOA depth attained the maximum HV for both transfer configurations and all random seeds. This indicates that the instance is sufficiently easy to solve, so that all tested parameter sets are capable of reaching the maximum HV. As the problem size increases, only a subset of QAOA depths reaches the maximum HV, and the time required to do so increases (see table \ref{tab:timecomparison}). Also, transferring parameters from training graphs closer in size to the target graph tends to produce better performance.

With respect to the QAOA depth, it is difficult to identify a general preference across the considered problems. For the TSP instances, this behavior is easier to analyze because the preferred depths remain relatively consistent across the different random seeds. For example, in the TSP-6 experiments, the maximum HV was reached fastest with $p=1$ when transferring parameters from 27-qubit training graphs, whereas $p=3$ performed best when transferring from 17-qubit graphs. However, the opposite trend is observed for TSP-5, where the fastest convergence is obtained with $p=3$ for the 17-qubit transfer and with $p=1$ for the 11-qubit transfer.

In contrast, the VRP experiments exhibit greater variability. Some parameter-transfer configurations fail to attain the maximum HV, and in a few cases, parameters transferred from smaller training graphs outperform those transferred from graphs closer in size to the target instance. For example, for the VRP-4-1 instance, neither the 14-qubit nor the 20-qubit transfer reaches the highest hypervolume observed across all experiments. This behavior was not restricted to a single instance, suggesting that transfer performance is not determined solely by the similarity in graph size.

The number of non-dominated solutions is relatively small for all considered instances, ranging from one to sixteen depending on the problem and random seed, in contrast to the much larger Pareto fronts reported for generic MAX-CUT instance by \cite{Kotil2025}. Regarding feasibility, almost all QAOA depths produce feasible solutions for the TSP. In contrast, the VRP exhibits a much stronger dependence on both the transferred parameters and the QAOA depth. Most transfer configurations yield feasible solutions only for a subset of depths, while some fail to produce any feasible solution for particular random seeds, as observed for the 20-qubit transfer in the VRP-5-1 instance.

Finally, we compare both approaches from a practical perspective by considering a fixed computational budget of 2,500 seconds. For the QAOA-based method, this corresponds to executing 25 million measurement shots, assuming a sampling rate of 10,000 shots per second. For the adapted $\epsilon$-constraint method, the same wall-clock budget was enforced by interrupting the optimization after 2,500 seconds on an Amazon EC2 instance.

\begin{figure}[!h]
    \centering
    \includegraphics[width=\linewidth]{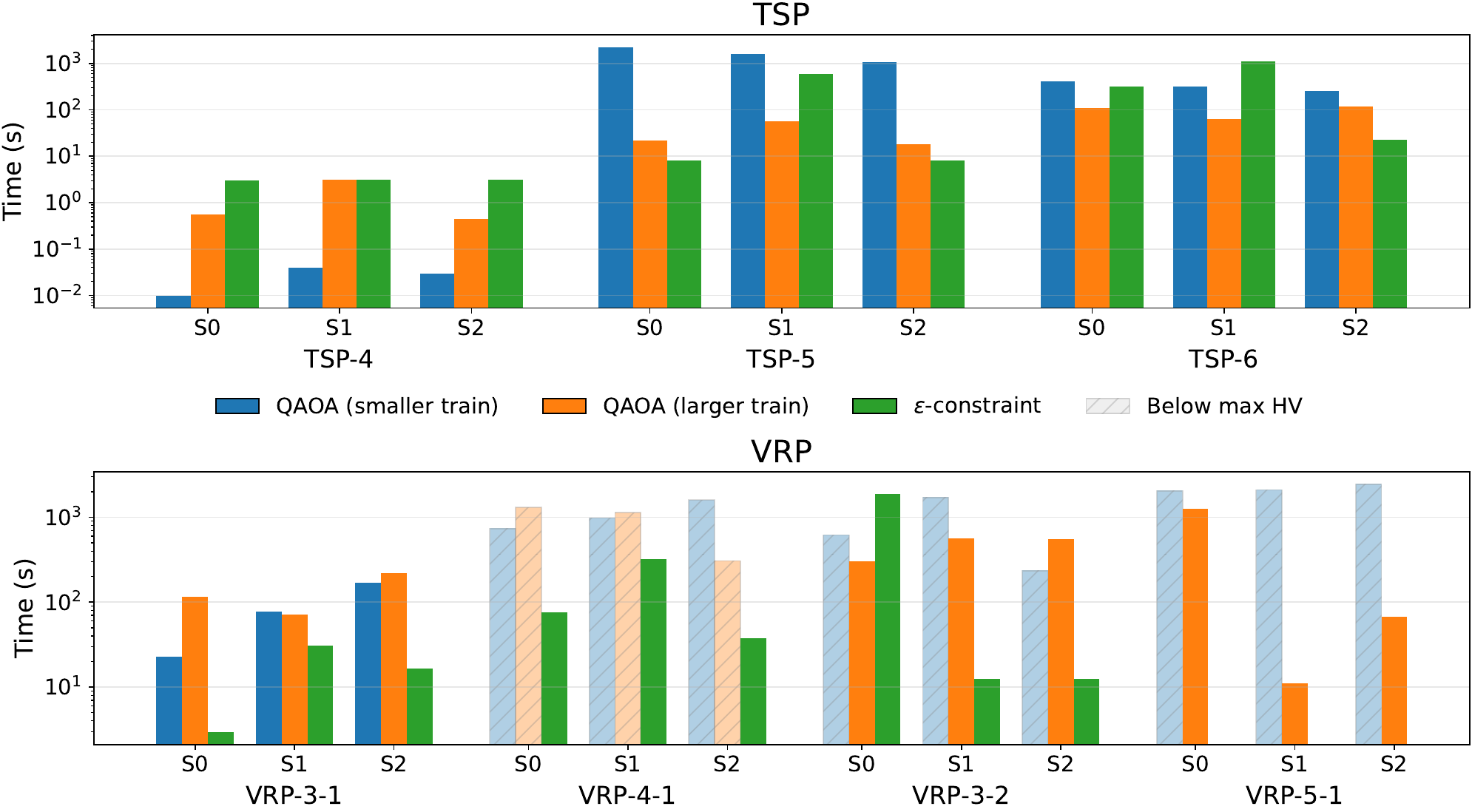}
    \caption{Execution time required to attain the highest hypervolume observed for each problem instance and random seed. The blue and orange bars correspond to QAOA using parameters transferred from the smaller and larger training graphs, respectively, while the green bars correspond to the adapted $\epsilon$-constraint method. Hatched bars indicate configurations that did not attain the highest hypervolume observed for the corresponding instance and random seed.}
    \label{fig:timecomparison}
\end{figure}

Figure~\ref{fig:timecomparison} summarizes the execution time required by each method to attain the highest hypervolume observed for each problem instance and random seed. The exact execution times and the corresponding QAOA depths are reported in Table~\ref{tab:timecomparison}.

\begin{table*}[!t]
\centering
\caption{Comparison between the fastest QAOA configuration reaching the maximum hypervolume and the adapted $\epsilon$-constraint method. For each seed, QAOA results are reported as execution time and the corresponding QAOA depth $p$. Entries marked with an asterisk (*) indicate that the corresponding QAOA configuration did not attain the highest hypervolume observed among all evaluated QAOA depths and the adapted $\epsilon$-constraint method for the same problem instance and random seed. Times shown in bold indicate that QAOA reached the maximum hypervolume faster than the adapted $\epsilon$-constraint method.}
\label{tab:timecomparison}
\begin{tabular}{ll|cc|cc|cc|ccc}
\hline
& &
\multicolumn{6}{c|}{QAOA} &
\multicolumn{3}{c}{$\epsilon$-constraint (s)}\\
\cline{3-11}
& &
\multicolumn{2}{c|}{Seed 0} &
\multicolumn{2}{c|}{Seed 1} &
\multicolumn{2}{c|}{Seed 2} &
\multirow{2}{*}{Seed 0} &
\multirow{2}{*}{Seed 1} &
\multirow{2}{*}{Seed 2}\\
\cline{3-8}
Instance & Train &
Time (s) & $p$ &
Time (s) & $p$ &
Time (s) & $p$ &
& & \\
\hline

\multirow{2}{*}{TSP-4}
& 7  
& \textbf{0.01} & 3 
& \textbf{0.04} & 1 
& \textbf{0.03} & 1
& \multirow{2}{*}{3.05}
& \multirow{2}{*}{3.16}
& \multirow{2}{*}{3.11}
\\
& 11 
& \textbf{0.56} & 3 
& 3.18 & 1 
& \textbf{0.44} & 1
& & &
\\
\hline

\multirow{2}{*}{TSP-5}
& 11  
& 2224.95 & 1 
& \textbf{1618.56} & 1 
& 1066.56 & 6
& \multirow{2}{*}{8.18}
& \multirow{2}{*}{595.93*}
& \multirow{2}{*}{8.07}
\\
& 17 
& 22.16 & 4 
& \textbf{56.89} & 4 
& 18.12 & 4
& & &
\\
\hline

\multirow{2}{*}{TSP-6}
& 17  
& \textbf{410.25} & 3 
& \textbf{323.16} & 3 
& \textbf{252.28} & 3
& \multirow{2}{*}{323.67*}
& \multirow{2}{*}{1123.84}
& \multirow{2}{*}{23.08*}
\\
& 27 
& \textbf{109.68} & 1 
& \textbf{64.23} & 1 
& \textbf{117.36} & 1
& & &
\\
\hline

\multirow{2}{*}{VRP-3-1}
& 8  
& 22.98 & 3 
& 77.90 & 1 
& 170.31 & 1
& \multirow{2}{*}{2.93}
& \multirow{2}{*}{31.05}
& \multirow{2}{*}{16.66}
\\
& 14 
& 115.96 & 2 
& 72.22 & 6 
& 220.24 & 6
& & &
\\
\hline

\multirow{2}{*}{VRP-4-1}
& 14  
& 733.09* & 2 
& 979.41* & 1 
& 1591.34* & 3
& \multirow{2}{*}{75.43}
& \multirow{2}{*}{323.59}
& \multirow{2}{*}{37.73}
\\
& 20 
& 1304.79* & 1 
& 1131.23* & 1 
& 305.79* & 1
& & &
\\
\hline

\multirow{2}{*}{VRP-3-2}
& 20  
& 618.51* & 2 
& 1700.44* & 3 
& 233.61* & 2
& \multirow{2}{*}{1876.46}
& \multirow{2}{*}{12.51*}
& \multirow{2}{*}{12.45}
\\
& 27 
& \textbf{300.83} & 6 
& \textbf{559.51} & 4 
& 548.84 & 4
& & &
\\
\hline

\multirow{2}{*}{VRP-5-1}
& 20  
& 2049.58* & 4 
& 2097.65* & 4 
& 2452.43* & 4
& \multirow{2}{*}{--}
& \multirow{2}{*}{--}
& \multirow{2}{*}{--}
\\
& 27 
& 1254.51 & 6 
& 11.14 & 5 
& 67.35 & 5
& & &
\\
\hline

\end{tabular}
\end{table*}

Several observations can be drawn from Figure~\ref{fig:timecomparison} and Table~\ref{tab:timecomparison}. First, the comparison only considers the time required to reach the highest hypervolume observed among both methods for each problem instance and random seed. Consequently, entries marked with an asterisk in Table~\ref{tab:timecomparison}, and the corresponding hatched bars in Figure~\ref{fig:timecomparison}, indicate cases in which the corresponding QAOA configuration did not attain this reference hypervolume within the available computational budget.

For the smallest instances, both approaches reach the highest hypervolume within only a few seconds, although QAOA is generally faster. As the problem size increases, the comparison becomes more heterogeneous. For the TSP instances, transferred QAOA parameters frequently allow the algorithm to attain the highest hypervolume substantially earlier than the adapted $\epsilon$-constraint method. This behavior is particularly evident for the TSP-6 experiments, where QAOA consistently reaches the reference hypervolume several hundred seconds before the classical approach whenever both methods succeed. The VRP instances present a more diverse behavior. In the VRP-4-1 experiments, the adapted $\epsilon$-constraint method consistently obtains a higher hypervolume than QAOA within the same computational budget.

Overall, the experimental results demonstrate that parameter transfer remains effective in the context of multi-objective combinatorial optimization beyond the MAX-CUT problems originally considered by \cite{Kotil2025}. For the TSP instances, transferred parameters consistently reached the highest hypervolume and frequently did so faster than the adapted $\epsilon$-constraint method. The VRP instances, however, exhibited considerably greater variability. While parameter transfer still produced competitive solutions and often reduced the time required to reach the best hypervolume, some transfer configurations failed to attain the highest hypervolume for a given instance and random seed.

The contrast between the two problem classes suggests that the effectiveness of parameter transfer depends on the structure of the optimization landscape. The TSP Hamiltonian appears to admit parameter regions that generalize well across graph sizes, whereas the additional decision variables and feasibility constraints introduced by the VRP lead to a more complex landscape, making successful parameter transfer less consistent.

\subsection{Hardware Validation}
To validate the parameter-transfer strategy beyond classical simulation, we executed the QAOA-based approach on IBM quantum hardware for the largest TSP and VRP instances considered in this work, specifically using the instances generated with Seed 0. We transferred parameters from 17-qubit training graphs to 37-qubit TSP instances (TSP-6), and from 27-qubit training graphs to 43-qubit VRP instances (VRP-5-1). The hardware experiments were conducted on the \textit{ibm\_kingston} quantum processor using 750 scalarization weight vectors and 5,000 measurement shots per scalarized instance, executed in batches of 50 samples per run cycle. This represents a restricted sampling budget corresponding to approximately $15\%$ of the samples used in our MPS simulations (5,000 samples). The QAOA depths were selected based on the best-performing configurations identified in the simulation study: $p = 3$ for the TSP instance and $p = 6$ for the VRP instance.

\begin{figure*}
\centering
\includegraphics[width=0.48\textwidth]{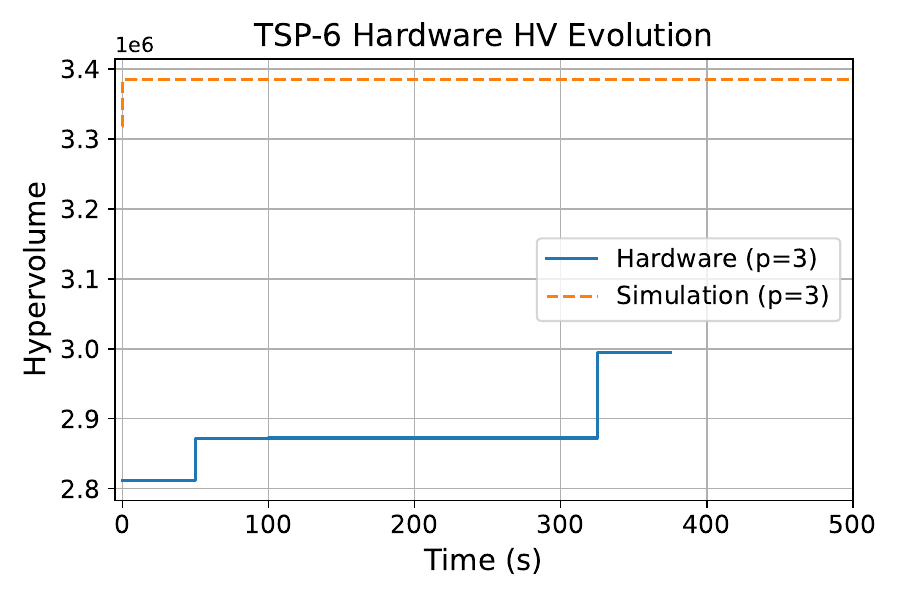}
\includegraphics[width=0.48\textwidth]{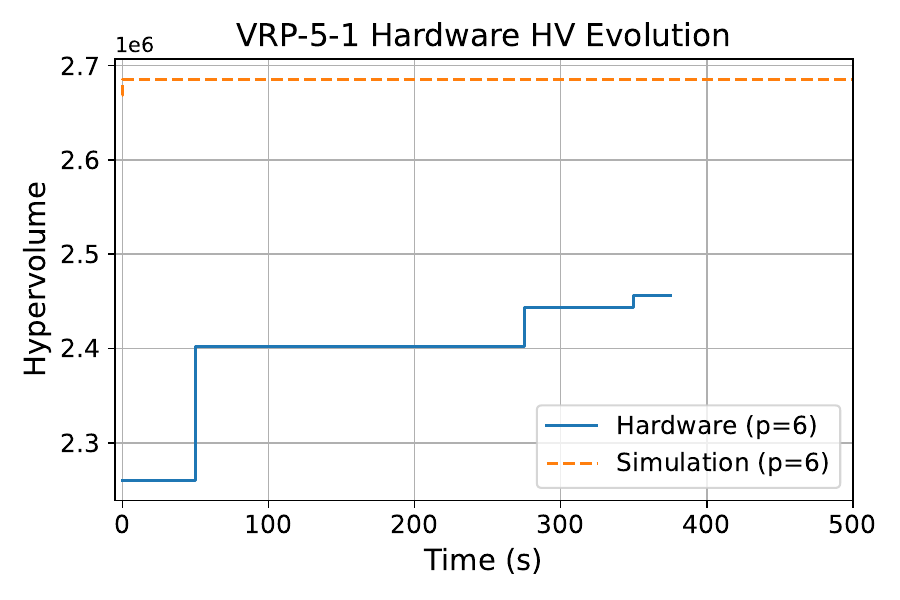}
\caption{Hypervolume evolution on IBM quantum hardware (solid lines) and noise-free MPS simulation (dashed lines) for TSP-6 (17$\to$37q, $p=3$, left) and VRP-5-1 (27$\to$43q, $p=6$, right). Only the QAOA depth tested on hardware is shown for each problem.}
\label{fig:hw_hv_evolution}
\end{figure*}

Figure~\ref{fig:hw_hv_evolution} compares the hypervolume evolution between the hardware and the noise-free MPS simulation for the corresponding QAOA depths ($p=3$ for TSP-6 and $p=6$ for VRP-5-1). For both experiments, the hardware did not find any feasible solution. As shown in Figure~\ref{fig:hw_hv_evolution}, there is a significant disparity between the hypervolume trajectories obtained on the hardware and those from the simulation, indicating that hardware noise substantially degrades solution quality. The x-axis is expressed as $(\text{samples} \times \text{shots})/10{,}000$ and limited to 500~s to provide context around the hardware budget of 375~s ($750 \times 5{,}000$ evaluated weights).

The inability of the hardware to produce feasible solutions can be directly attributed to the extreme circuit depths required by the QAOA ansatz for these problem sizes. After transpilation to the native gate set of the \textit{ibm\_kingston} processor (a Heron-class device with 156 physical qubits), the TSP circuit ($p=3$, 37 logical qubits) has a transpiled depth of 6{,}328 and requires 11{,}707 two-qubit CZ gates, while the VRP circuit ($p=6$, 43 logical qubits) reaches a transpiled depth of 16{,}357 with 31{,}794 two-qubit CZ gates. By tracking the critical path of the scheduled operations based on typical Heron gate durations ($\approx$ 200~ns for CZ and $\approx$ 35.5~ns for single-qubit gates), the exact circuit execution times are 599~$\mu$s for the TSP and 1{,}550~$\mu$s for the VRP. These values far exceed the typical coherence times of the \textit{ibm\_kingston} qubits, which are on the order of $T_1 \approx T_2 \approx 200$~$\mu$s. Consequently, the TSP circuit runs approximately $3\times$ longer and the VRP circuit almost $8\times$ longer than the qubit coherence window, leading to a massive accumulation of decoherence and gate errors that prevents the algorithm from satisfying the hard constraints of the routing problems.

Given these physical limitations, successfully deploying QAOA for multi-objective routing problems on near-term hardware will require either substantial improvements in qubit coherence and gate fidelities or the development of more efficient circuit compilation and error mitigation strategies.

\section{Conclusion and Future Works}
\label{sec:conclusion}

In this paper, we evaluated the effectiveness of the QAOA parameter transfer strategy for multi-objective optimization proposed by \cite{Kotil2025}. The original work showed that MOO problems can be effectively solved using QAOA within a weighted-sum method (WSM) by evaluating a large set of scalarization weights. Its main contribution was the use of parameter transfer, where QAOA parameters are learned on a smaller problem and then transferred to a larger one for each scalarization weight. The authors demonstrated the effectiveness of this strategy on the MAX-CUT problem using MPS simulations and real quantum hardware.

Since their evaluation scenario is somewhat idealized, as MAX-CUT has no direct practical application and the graphs used follow the topology of the target quantum hardware, we performed a comprehensive benchmark of the method on the TSP and VRP. Both problems were formulated as QUBO models, reduced to MAX-CUT, and evaluated under the same experimental conditions adopted in the original paper. We also validated the parameter transfer strategy on real IBM quantum hardware.

Our simulation results show that the parameter transfer strategy also works successfully for TSP and VRP, often producing higher hypervolume values than the adapted $\epsilon$-constraint method while requiring less computational time. However, hardware experiments on the \textit{ibm\_kingston} processor did not produce any feasible solution for either problem, due to the extreme circuit depths required (transpiled depths of 6{,}328 and 16{,}357 for the TSP and VRP, respectively), which far exceed the coherence times of the device. This behavior differs substantially from that reported for MAX-CUT. The best QAOA depth varies considerably depending on the problem size, the size of the training instance, and, in some cases, the random seed. Moreover, parameter transfer is considerably more stable for the TSP than for the VRP.

The effectiveness of parameter transfer appears to depend strongly on the underlying optimization problem. Since the TSP and, particularly, the VRP lead to more complex and highly constrained Hamiltonians than MAX-CUT, the corresponding optimal QAOA parameters appear to be less concentrated, making successful transfer more challenging.

Finally, several aspects of parameter transfer in QAOA remain open for further investigation. Although some theoretical studies have recently explored this phenomenon, extensive experimental evaluations are still required. In particular, benchmarking parameter-transfer strategies across a broader range of optimization problems, from simple to highly constrained formulations, especially in the single-objective setting, may provide further insights into the conditions under which parameter transfer is effective and the factors that influence its performance.

\section*{Acknowledgments}
This work was executed under the TIC26 – Brazil Quantum Camp project, funded within the scope of the Prioritized Informatics Programs and Projects (PPI), Process No. 01245.008254/2025-22, under the responsibility of the Ministry of Science, Technology and Innovation (MCTI), with operational coordination by the Association for the Promotion of Brazilian Software Excellence (SOFTEX), and executed by CESAR and the Instituto de Pesquisas ELDORADO.

%\funding{Sample text inserted for demonstration.}
% This section is a list of funder names and grant numbers

%\roles{Sample text inserted for demonstration.}
% List author names and the contributions made to the article, using terms from the NISO Contributor Roles Taxonomy (CRediT) https://credit.niso.org

%\data{Sample text inserted for demonstration.}
% For more information on IOP Publishing's research data policy see: https://publishingsupport.iopscience.iop.org/questions/research-data/

%\suppdata{Sample text inserted for demonstration.}

\bibliographystyle{unsrt}
\bibliography{bibliography}

\end{document}